\documentclass[%
reprint,
superscriptaddress,
amsmath,amssymb,
aps,
]{revtex4-2}

\usepackage{graphicx}% Include figure files
\usepackage{dcolumn}% Align table columns on decimal point
\usepackage{bm}% bold math
\usepackage{mathrsfs}
\usepackage{float}
\usepackage{physics}
\usepackage{siunitx}
\usepackage[colorlinks,linkcolor=blue,anchorcolor=blue,urlcolor=blue,citecolor=blue]{hyperref}
\begin{document}
%\linenumbers % 开启行号

\preprint{APS/123-QED}

\title{A Solvable Semi-infinite Fock-state-lattice SSH Model: the Stable Topological Zero Mode and the Non-Hermitian Bound Effect}

\author{Xing Yao Mi}
\affiliation{%
	Institute for Quantum Science and Technology, College of Science, National University of Defense Technology, Changsha, Hunan 410073, China\\
}% 
\affiliation{%
	Hunan Key Laboratory of Mechanism and Technology of Quantum Information, Changsha, Hunan 410073, China\\
}%

\author{Yong-Chun Liu}
\affiliation{%
	State Key Laboratory of Low-Dimensional Quantum Physics, Department of Physics, Frontier Science Center for Quantum Information, Tsinghua University, Beijing 100084, China\\
}% 

\author{Zhi Jiao Deng}%
\email{dengzhijiao926@hotmail.com}
\affiliation{%
	Institute for Quantum Science and Technology, College of Science, National University of Defense Technology, Changsha, Hunan 410073, China\\
}% 
\affiliation{%
	Hunan Key Laboratory of Mechanism and Technology of Quantum Information, Changsha, Hunan 410073, China\\
}%

\affiliation{%
    Key Laboratory of Low Dimensional Quantum Structures and Quantum Control of Ministry of Education, Hunan Normal University, Changsha, Hunan, 410081, China
}%

\author{Chun Wang Wu}%
\email{cwwu@nudt.edu.cn}
\affiliation{%
	Institute for Quantum Science and Technology, College of Science, National University of Defense Technology, Changsha, Hunan 410073, China\\
}% 
\affiliation{%
	Hunan Key Laboratory of Mechanism and Technology of Quantum Information, Changsha, Hunan 410073, China\\
}% 

\author{Ping Xing Chen}%
\affiliation{%
	Institute for Quantum Science and Technology, College of Science, National University of Defense Technology, Changsha, Hunan 410073, China\\
}% 
\affiliation{%
	Hunan Key Laboratory of Mechanism and Technology of Quantum Information, Changsha, Hunan 410073, China\\
}% 

\date{\today}% It is always \today, today,

\begin{abstract}
Fock-state lattice (FSL) offers a powerful quantum simulator for topological phenomena due to the unbounded scalability and ease of implementation. Nevertheless, the unique topological properties induced by its site-dependent coupling have remained elusive, mainly due to the challenge of handling an infinite state space without translational symmetry. Here, we rigorously analyze the topological features of a semi-infinite FSL-based Su-Schrieffer-Heeger (SSH) model, in both Hermitian and non-Hermitian realms, by mapping it to the solvable Jaynes-Cummings (JC) model via a unitary displacement transformation. We find a topological zero mode persisting across all parameter regimes, which is more stable than the conventional SSH model. It originates from the bound state at the inherent domain wall under anisotropic conditions. With gain and loss introduced, we predict a non-Hermitian bound effect (NHBE), i.e., any state overlapping with the bound state will quickly stabilize to the domain wall, with the minimal stabilization time occurring in the vicinity of  exceptional point (EP). The parity-time ($\mathcal{PT}$) phase transition can be observed by the oscillating-to-steady crossover of dynamics in the subspace orthogonal to the bound state. Furthermore, a concrete experimental proposal based on the trapped-ion setup is provided. 
\end{abstract}

\maketitle

\textit{Introduction.}$-$Since the seminal discovery of the quantum Hall effect \cite{PhysRevLett.45.494, PhysRevLett.49.405, PhysRevB.23.5632, RevModPhys.82.3045, RevModPhys.83.1057}, topological phenomena have become research hotspots in physics and are now expanding into non-Hermitian realms.
Non-Hermitian topological systems, characterized by complex spectra \cite{PhysRevX.8.031079, PhysRevLett.120.146402, Ashida2020, PhysRevLett.116.133903, PhysRevLett.121.026808, Xiao2020}, possess unique properties from their Hermitian counterparts, e.\,g.\,non-Bloch bands \cite{PhysRevLett.121.086803, PhysRevX.9.041015, PhysRevLett.121.136802, PhysRevLett.123.066404, RevModPhys.93.015005}, generalized Brillouin zones \cite{PhysRevLett.121.086803, science.aap9859, PhysRevLett.125.126402, PhysRevLett.124.086801}, and the non-Hermitian skin effect \cite{PhysRevLett.121.086803, PhysRevLett.128.223903, PhysRevLett.123.246801, Helbig2020, PhysRevLett.123.016805, PhysRevLett.123.170401}. On the other side, observing topological phenomena in practical condensed matter is a great challenge, which stimulates simulations of topological models in artificial quantum systems \cite{Bloch2012, Dalibard2011, Gross2017}, but still limited by small manipulation scales. The recently developed synthetic dimensions \cite{Boada2012, Ozawa2019, Celi2014, RevModPhys.97.011001} provide a solution to this limitation by utilizing the degrees of freedom in  physical systems as real spatial dimensions, which can be extended to any high dimension and implemented on diverse platforms like circuit networks \cite{Helbig2020, Imhof2018, Lee2018}, photons \cite{Yuan2018, Bandres2018, Lu2014, Ozawa2019b, Lustig2019, Hafezi2011, Maczewsky2020, Dutt2020}, optical lattices \cite{Aidelsburger2013, Miyake2013, Price2017, Tarruell2012}, and atomic momentum lattices \cite{Gadway2015, Meier2016, An2017, Li2022, Liang2024, Li2023, Liang2022b}.

Fock-state lattice (FSL) \cite{Saugmann2023, Wang2024, Oliver2023, PhysRevLett.134.070601, PerezLeija2010, PerezLeija2016, Keil2011, Wang2016, Cai2020, Wu2023, Yuan2024, Deng2022, Yang2024} emerges as a leading synthetic-dimension platform, hosting an intrinsic infinite-dimensional Hilbert space comprised of Fock states. However, its  inhomogeneous coupling strength ($\propto\sqrt{n}$, with $n$ the excitation number) breaks the translational symmetry of the lattice and poses a challenge for analyzing infinite-dimensional models based on FSLs \cite{Price2017, Saugmann2023, Wang2024, Oliver2023, PhysRevLett.134.070601, Wang2016, Cai2020, Wu2023, Yuan2024, Deng2022, Yang2024}. Whether and what distinct topological properties, compared to their isotropic counterparts, can be induced by this $\sqrt{n}$-dependent coupling, is an urgent question to answer. We note that, previous studies on FSLs have either focused on topological models that are robust to inhomogeneous coupling \cite{Price2017, Wang2024}, or limited to considering a finite-dimensional subspace with a fixed excitation number in multiple resonators \cite{Wang2016, Cai2020, Wu2023, Yuan2024, Deng2022, Yang2024}. The aforementioned question still remains unresolved.

In this work, we systematically investigate both Hermitian and non-Hermitian topological properties in a semi-infinite FSL-based Su-Schrieffer-Heeger (SSH) chain. By establishing a connection to the solvable Jaynes-Cummings (JC) model through displacement transformation, we demonstrate two features absent in conventional SSH models: a more stable topological zero mode and a non-Hermitian bound effect (NHBE, i.\,e., tendency of quickly stabilizing to the bound state), both originating from the site-dependent-coupling-induced topological domain wall. We also identify two signatures of the exceptional point (EP) in the parity-time ($\mathcal{PT}$)-symmetry-breaking transition \cite{PhysRevLett.103.093902, ElGanainy2018}, one being the minimal stabilization time for NHBE, and the other being the transition point of oscillating-to-steady behaviors. Finally, we propose a feasible experimental scheme based on the trapped-ion platform. Our study offers an analytical framework for investigation of infinite-dimensional anisotropic topological models and may inspire further experimental exploration of topological properties on FSL platforms.

\begin{figure}[t]
	\centering
	\includegraphics[width=8.6cm,height=4.3cm]{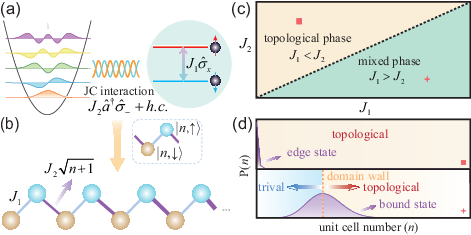}% Here is how to import EPS art
	\caption{\label{fig1}  A semi-infinite anisotropic SSH model and its topological phase diagram. [(a), (b)] A spin-boson system with JC coupling and a resonant spin driving, can be mapped onto a SSH chain, with intracell coupling $J_1$, anisotropic intercell coupling $J_2\left( n \right) =J_2\sqrt{n+1}$, and the $n$-th unit cell $\{\ket{n,\uparrow},\ket{n,\downarrow }\}$. [(c), (d)] The anisotropic SSH model exhibits a topological phase for $J_1 < J_2$, hosting a single topological edge state at the left boundary, and a mixed phase for $J_1 > J_2$, characterized by a topological domain wall separating the trivial and topological region, with a bound state localized at the interface.}
\end{figure}

\textit{Basic model.}$-$The model we consider is a spin-boson system with JC coupling and a resonant spin driving [Fig.~\ref{fig1}(a)]. The Hamiltonian is given by
\begin{equation}
	\begin{aligned}
		\hat{H}&= J_1\sum_{n=0}^{\infty}{\left( \ket{ n,\uparrow} \bra{n,\downarrow}+\mathrm{h.c.} \right)}\\ 
		&\quad+J_2\sum_{n=0}^{\infty}{\sqrt{n+1}\left( \ket{ n+1,\downarrow } \bra{n,\uparrow}+\mathrm{h.c.} \right)}\\ 
		&=J_1\hat{\sigma}_x+J_2\left( \hat{a}^{\dag}\hat{\sigma}_-+\hat{a}\hat{\sigma}_+ \right),
	\end{aligned}
	\label{eq:one}
\end{equation}
where $\ket{\uparrow \left( \downarrow \right)}$ represents the spin-up (down) state, $J_1$ and $J_2$ are interaction strengths of the spin driving and the JC coupling, $\hat{a}$ and $\hat{a}^{\dag}$ are the bosonic annihilation and creation operators on the Fock-state basis $\left| n \right>$, $\hat{\sigma}_{x,y,z}$ are the spin Pauli operators, and $\hat{\sigma}_{\pm}=\left( \hat{\sigma}_x\pm i\hat{\sigma}_y \right) /2$ are the spin raising and lowering operators. By reinterpreting the bosonic Fock-state space as a synthetic spatial dimension and treating spin-boson states $\{\ket{ n,\uparrow},\ket{ n,\downarrow }\}$ as the $n$-th unit cell, the above driven-JC system can be mapped onto a SSH chain with intracell coupling $J_1$ and anisotropic intercell coupling $J_2\left( n \right) =J_2\sqrt{n+1}$ [Fig.~\ref{fig1}(b)]. Experimentally, this model admits direct implementation on diverse quantum optical platforms including nitrogen-vacancy (NV) centers \cite{Teissier2014, Barfuss2015}, trapped ions \cite{Kienzler2015}, atoms in optical lattices \cite{Aidelsburger2013, Miyake2013, An2017, Li2022}, nano-mechanical resonators (NAMR) \cite{Chan2011, Teufel2011}, and cavity quantum electrodynamics (CQED) \cite{Kerckhoff2009, Mabuchi2002, Boissonneault2008, Hamsen2017}.

The site-dependent coupling in this anisotropic SSH model induces a distinctive topological phase diagram [Fig.~\ref{fig1}(c, d)]. For $J_1 < J_2$, the SSH chain resides in the topological phase, and owing to its semi-infinite geometry, hosts a single topological edge state localized at the left boundary. For $J_1 > J_2$, it separates into a topologically trivial region [$J_1 > J_2(n)$] and a topological region [$J_1 < J_2(n)$] through a domain-wall interface, giving rise to a mixed phase absent in isotropic counterparts. As demonstrated in subsequent analysis, this site-dependent-coupling-induced domain wall yields unique topological features in the anisotropic SSH model, including a more stable zero mode than the isotropic case, corresponding to the bound state \cite{lajci2025} at the domain wall, and a non-Hermiticity enhanced bound effect.

\begin{figure}[b]
	\includegraphics[width=8.6cm,height=4.9235cm]{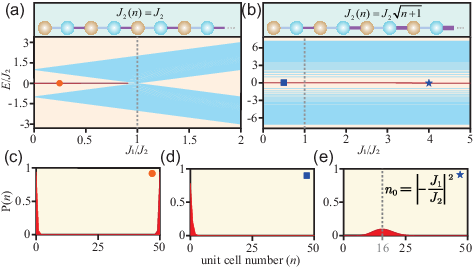}% Here is how to import EPS art
	\caption{\label{fig2} Comparison of the eigenenergies between isotropic and anisotropic SSH models. (a) The eigenenergy spectrum of the conventional isotropic SSH model is numerically obtained under open boundary conditions with unit-cell number truncated at $n=50$. It exhibits topological zero modes in the topological phase ($J_1<J_2$). (b) The eigenenergy spectrum of the anisotropic SSH model is plotted using the analytical eigenvalues in Eq.~(\ref{eq:eigenenergy}) with $n \leqslant 50$, showing a stable topological zero mode regardless of $J_1/J_2$. (c) Left and right edge states corresponding to the two-fold zero modes at $J_1/J_2=0.25$ (orange circle) in (a). [(d), (e)] The left-edge-localized state (d) and the domain-wall-bounded state (e), are plotted by using the analytical expression of zero modes $\ket{\alpha,\downarrow}$ at $J_1/J_2=0.5$ (blue square) and $J_1/J_2=4$ (blue pentagram) in (b), respectively.}
\end{figure}

\begin{figure*}[t]
	\includegraphics{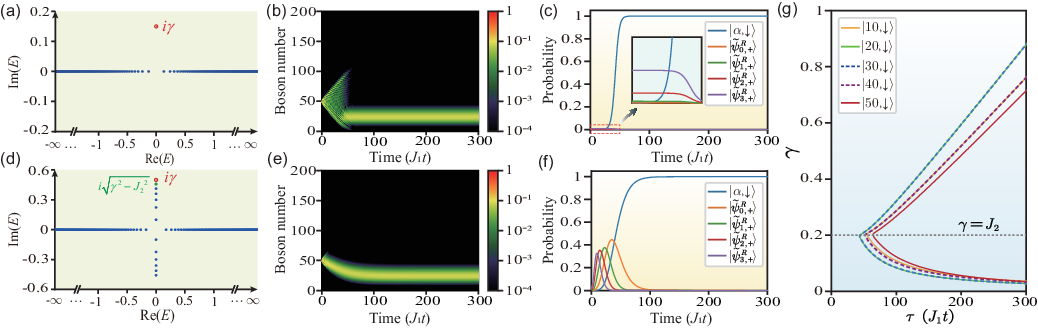}% Here is how to import EPS art
	\caption{\label{fig3}Non-Hermitian bound effect (NHBE) in the NH anisotropic SSH model. [(a)-(c)] In the $J_2 > \gamma$ parameter regime, the complex eigenspectrum hosts a single gain mode corresponding to the bound state $\ket{\varphi_0}$ (a). Both the boson number distribution (b) and projection probabilities onto representative eigenmodes (c) demonstrate that the renormalized time-evolved state is rapidly dominated by the exponentially amplified $\ket{\varphi_0}$, ultimately stabilizing at the domain-wall position. [(d)-(f)] In the $J_2 < \gamma$ parameter regime, additional gain modes besides the bound state emerge with  $|\mathrm{Im}(E)| < \gamma$ (d). The boson number distribution (e) and eigenmode projections (f) show that $\ket{\varphi_0}$ sustains its dominance in mode competition dynamics, eventually giving rise to NHBE. (g) Stabilization time $\tau$ versus $\gamma$ for representative initial states $\ket{n, \downarrow}$ ($n=10, 20, 30, 40, 50$) reveals that the fastest stabilization is near $\gamma = J_2$. Parameters: $J_1=1$, $J_2=0.2$; $\gamma=0.15$ [(a)-(c)], $\gamma=0.5$ [(d)-(f)]; the initial state is chosen as $\ket{ 50,\downarrow}$ [(b), (c), (e), (f)].	 
	}
\end{figure*}

\textit{The stable topological zero mode.}$-$Defining the chiral operator as $\mathcal{C}=\hat{\sigma}_z$, the Hamiltonian in Eq.~(\ref{eq:one}) satisfies the chiral symmetry $\mathcal{C}\hat{H}\mathcal{C}^{\dag}=-\hat{H}$, which indicates a symmetric spectrum about zero energy and may induce topologically protected zero modes when the energy gap opens. Note that, directly diagonalizing $\hat{H}$ to obtain the exact eigenmodes in the Fock-state basis is challenging, as its infinite-dimensional Hilbert space cannot be decomposed into a direct sum of decoupled finite-dimensional subspaces. Additionally, the site-dependent coupling explicitly breaks the lattice's translational symmetry, thereby excluding Bloch states as valid eigenstate candidates. However, these difficulties can be elegantly circumvented by establishing an exact mapping between $\hat{H}$ and the solvable JC-Model $\hat{H}_{JC}=J_2\left(\hat{a}^{\dag}\hat{\sigma}_-+\hat{a}\hat{\sigma}_+ \right)$ via an appropriate unitary transformation (see Appendix~\ref{AppendixA}), i.\,e.,
\begin{equation}
	\hat{D}^{\dagger}(\alpha) \hat{H} \hat{D}(\alpha) = \hat{H}_{JC},
\end{equation}
where $\hat{D}(\alpha) = e^{\alpha \hat{a}^{\dagger} - \alpha^* \hat{a}}$ is the displacement operator with $\alpha = -J_1/J_2$. Then, using the displaced-Fock-state basis, we can obtain the exact eigenstates of $\hat{H}$ as
\begin{equation}
\begin{aligned}
	\label{eq:eigenstate}
	\ket{\varphi_0} &=\hat{D}(\alpha)\ket{0, \downarrow}=\ket{\alpha, \downarrow}, \\
	\ket{\varphi_{n, \pm}} &= \hat{D}(\alpha) (\ket{n, \uparrow} \pm \ket{n+1, \downarrow})/\sqrt{2},
\end{aligned}
\end{equation}
and the corresponding eigenvalues as
\begin{equation}
	E = 0, \pm J_2 \sqrt{n+1},
	\label{eq:eigenenergy}
\end{equation}
where $n = 0, 1, 2, \cdots, \infty$. Eqs.~(\ref{eq:eigenstate},\ref{eq:eigenenergy}) reveal that the anisotropic SSH model exhibits a more stable zero mode than the isotropic case: the zero mode persists across all parameter regimes of $J_1/J_2$, whereas the isotropic model sustains zero modes exclusively when $J_1 < J_2$.

To illustrate this, Fig.~\ref{fig2} compares eigenenergies as functions of $J_1/J_2$ between the isotropic SSH model and our anisotropic counterpart. It shows that the isotropic model exhibits twofold-degenerate zero modes exclusively in the topological phase ($J_1 < J_2$), associated with left and right edge states [Fig.~\ref{fig2}(a, c)]. In contrast, the anisotropic model hosts a stable zero mode regardless of the value of $J_1/J_2$, manifesting as a left-edge-localized state in the topological phase ($J_1 < J_2$) or a domain-wall-bounded state centered at $n_0=|\alpha|^2$ in the mixed phase ($J_1 > J_2$) [Fig.~\ref{fig2}(b, d, e)]. The discrepancy between the topological boundary in Fig.~\ref{fig2}(a) and $J_1/J_2=1$ is due to the finite-size effect, which also leads to a sensitive dependence of the edge states on the system size (a situation in finite topological models where the two edge states hybridize when they have spatial overlap). In contrast, our semi-infinite FSL-based anisotropic SSH chain inherently hosts a single zero mode, thus is unaffected by this effect.

\textit{The non-Hermitian bound effect (NHBE).}$-$In our model, introducing balanced gain $i\gamma$ (on $\ket \downarrow$) and loss $-i\gamma$ (on $\ket\uparrow$) yields a anisotropic non-Hermitian (NH) SSH chain, with the Hamiltonian given by
\begin{equation}
	\hat{H}_\mathrm{NH} = J_1 \hat{\sigma}_x + J_2 (\hat{a}^\dagger \hat{\sigma}_- + \hat{a} \hat{\sigma}_+) - i\gamma \hat{\sigma}_z.
	\label{eq:NHHamiltonian}
\end{equation}
Following the same procedure as the Hermitian case, we obtain the analytically exact eigenenergies of $\hat{H}_\mathrm{NH}$ as
\begin{equation}
	E_0 = i\gamma, E_{n, \pm} = \pm \sqrt{J_2^2 (n+1) - \gamma^2}
	\label{eq:NHeigenvalues}
\end{equation} 
and the corresponding eigenstates as
\begin{equation}
	\begin{aligned}
		\ket{\varphi_0} &= \ket{\alpha, \downarrow},\\
		\ket{ \psi _{n,\pm} } &= J_2\sqrt{n+1}\ket{ \varphi _1(n) }\\ 
		&\quad+\left[ i\gamma \pm \sqrt{J_{2}^{2}(n+1) -\gamma ^2} \right] \ket{ \varphi _2(n) }, 
	\end{aligned}
	\label{eq:NHeigenstates}
\end{equation}
where $\ket{\varphi_1(n)} = \hat{D}(\alpha) \ket{n, \uparrow}$ and $\ket{\varphi_2(n)} = \hat{D}(\alpha) \ket{n+1, \downarrow}$ (with $\alpha = -J_1/J_2$ and $n = 0, 1, 2, \cdots, \infty$) are displaced Fock states. Employing the biorthogonal completeness relation
\begin{equation}
	\ket{\alpha ,\downarrow} \bra{\alpha ,\downarrow}+\sum_{n=0}^{\infty}{\sum_{g=+,-}{\ket{ \tilde{\psi}_{n,g}^{R}} \bra{\tilde{\psi}_{n,g}^{L} }}=\hat{I}},
	\label{eq:completerelation}
\end{equation}
where biorthogonal normalized right (left) eigenstates $\ket{ \tilde{\psi}_{n, g}^{R(L)} }$ are the biorthogonal basis for $\hat{H}_\mathrm{NH}$ satisfying $\hat{H}_\mathrm{NH}\ket{ \tilde{\psi}_{n, g}^{R} }=E_{n, g}\ket{ \tilde{\psi}_{n, g}^{R} }$,  $\hat{H}_\mathrm{NH}^{\dagger}\ket{ \tilde{\psi}_{n, g}^{L} }=E_{n, g}^{*}\ket{ \tilde{\psi}_{n, g}^{L} }$, and $\braket{\tilde{\psi}_{n,g}^{L}}{\tilde{\psi}_{n^{\prime}, g^{\prime}}^{R}} = \delta_{nn^{\prime}}\delta_{gg^{\prime}}$, the time-evolved state takes the form (see Appendix~\ref{AppendixC})
\begin{equation}
	\ket{\varPsi(t)} =c_0e^{\gamma t}\ket{\alpha ,\downarrow} +\sum_{n=0}^{\infty}{\sum_{g=+,-}{c_{n,g}e^{-iE_{n,g}t}\ket{ \tilde{\psi}_{n,g}^{R} }}}
	\label{eq:evolution}
\end{equation}
with $c_0=\braket{\alpha,\downarrow}{\varPsi(0)}$ and $c_{n,g}=\braket{\tilde{\psi}_{n,g}^{L}}{\varPsi(0)}$ denoting the biorthogonal expansion coefficients of the initial state $\ket{\varPsi(0)}$. Eqs.~(\ref{eq:NHeigenvalues}--\ref{eq:evolution}) reveal that the topological bound state $\ket{\varphi_0}$ persists as the largest-gain mode independent of $J_1$, $J_2$ and $\gamma$ ($\neq0$), indicating a distinct non-Hermiticity induced bound effect: any initial state overlapping with the bound state exponentially stabilizes toward the domain wall.

\begin{figure}[t]
	\includegraphics{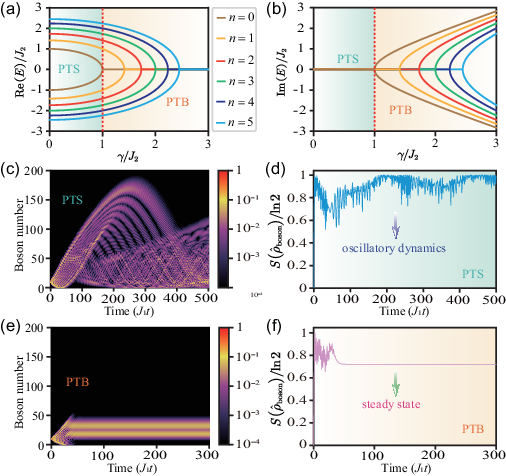}% Here is how to import EPS art
	\caption{\label{fig4} $\mathcal{P}\mathcal{T}$ quantum phase transition observed in the restricted Hilbert space governed by $\hat{H}_1$. [(a), (b)] Real (a) and imaginary (b) components of eigenvalues of $\hat{H}_1$. [(c), (d)] In the PTS phase, both boson-number dynamics (c) and spin-boson entanglement (d) exhibit oscillatory behaviors, with entanglement quantified by the von Neumann entropy $S(\hat{\rho}_\mathrm{boson})$. [(e), (f)] In the PTB phase, boson number dynamics (e) and entanglement entropy (f) rapidly converge to the values determined by the largest-gain mode $\ket{\tilde{\psi}_{0,+}^R}$ in $\hat{H}_1$, as marked by the green dot in Fig.~\ref{fig3}(d), after transient mode competition dynamics. Parameters: $J_1=1$, $J_2=0.2$; $\gamma=0.15$ [(c), (d)], $\gamma=0.25$[(e), (f)]; the initial state is chosen as $\ket{10,\uparrow}$ [(c)-(f)].}
\end{figure}

To verify NHBE, we analyze the eigenenergy spectrum and dynamical behavior of the NH anisotropic SSH chain under parameter regimes of $J_2 > \gamma$ and $J_2 < \gamma$, respectively [Fig.~\ref{fig3}]. In the former case, all eigenmodes except the bound state are conservative modes with $\mathrm{Im} (E)=0$ [Fig.~\ref{fig3}(a)]; consequently, the time-evolved state $\ket{ \varPsi(t) }$ is rapidly dominated by the exponentially amplified mode $\ket{\varphi_0}$, eventually localized at the domain wall [Fig.~\ref{fig3}(b)]. In the latter case, a subset of eigenvalues appear as complex-conjugate pairs with $|\mathrm{Im}(E)| < \gamma$ [Fig.~\ref{fig3}(d)], where the bound state still prevails in mode competition against other gain modes in long-time dynamics, thereby demonstrating the NHBE [Fig.~\ref{fig3}(e)]. The temporal evolution of projection probabilities for the renormalized time-evolved state onto the bound state and other representative eigenmodes reveals mode competition dynamics in the $J_2 > \gamma$ [Fig.~\ref{fig3}(c)] and $J_2 < \gamma$ [Fig.~\ref{fig3}(f)] regimes, respectively. It should be noted that, different renormalization procedures must be employed to obtain the projection probabilities for orthogonal boson number states [Fig.~\ref{fig3}(b, e)] and non-orthogonal eigenmodes [Fig.~\ref{fig3}(c, f)], respectively (see Appendix~\ref{AppendixC}). The relative gain factor of the bound state versus the dominant competing eigenmode in long-time dynamics corresponds to their imaginary energy gap $|\mathrm{Im}(\Delta E)|$, which monotonically increases with $\gamma$ for $\gamma < J_2$, decreases with $\gamma$ for $\gamma > J_2$, and attains its maximum at $\gamma = J_2$. Consequently, the required stabilization time $\tau$ of NHBE reaches its minimum near $\gamma = J_2$. In Fig.~\ref{fig3}(g), the plotted $\tau$ as a function of $\gamma$ for representative initial states $\ket{10, \downarrow}$, $\ket{20, \downarrow}$, $\ket{30, \downarrow}$, $\ket{40, \downarrow}$, and $\ket{50, \downarrow}$ provides strong validation for this conjecture. The observed deviations between minimal-time positions and $\gamma = J_2$ arise from state-dependent transient mode competition dynamics. As demonstrated in the following, the minimal stabilization time attained at $\gamma \approx J_2$ can actually serve as a distinguishing signature for the EP constructed within a restricted Hilbert space excluding the bound state.

\textit{Observation of $\mathcal{PT}$ quantum phase transition.}$-$It follows from Eqs.\,(\ref{eq:NHHamiltonian}--\ref{eq:NHeigenstates}) that, employing the displaced Fock states $\{\ket{\varphi_0}, \ket{\varphi_1(n)}, \ket{\varphi_2(n)} (n = 0, 1, 2, \cdots, \infty)\}$ as basis vectors, $\hat{H}_\mathrm{NH}$ can be decomposed into a direct sum of subspace Hamiltonians: 
\begin{equation}
	\begin{aligned}
		\label{eq:Hdirectsum}
		\hat{H}_\mathrm{NH} &= \hat{H}_0 \oplus \hat{H}_1,\\  
		\hat{H}_0 &= i\gamma \ket{\varphi_0}\langle\varphi_0|,\\ 
		\hat{H}_1 &= \bigoplus_{n=0}^{\infty} \left[J_2 \sqrt{n+1} \hat{X}_n - i\gamma \hat{Z}_n\right],
	\end{aligned}
\end{equation}
where $\hat{X}_n$ and $\hat{Z}_n$ represent Pauli $x$- and $z$-operators in the subspace $\{\ket{\varphi_1(n)}, \ket{\varphi_2(n)}\}$. By defining the generalized parity operator $\hat{\mathcal{P}}\equiv\bigoplus_{n=0}^{\infty}\hat{X}_n$ and the time-reversal operator $\hat{\mathcal{T}}$ as complex conjugation \cite{Huang2018,Huang2018b}, one can readily verify that
\begin{equation}
	(\hat{\mathcal{P}}\hat{\mathcal{T}}) \hat{H}_1 (\hat{\mathcal{P}}\hat{\mathcal{T}})^{-1} = \hat{H}_1,
	\label{eq:PTsymmetry}
\end{equation}
indicating the existence of $\mathcal{PT}$-symmetry-breaking quantum phase transition within the restricted Hilbert space $\{\ket{\varphi_1(n)}, \ket{\varphi_2(n)} (n = 0, 1, 2, \cdots, \infty)\}$ \cite{Supplementary1}. As shown in Fig.\,\ref{fig4}(a, b), for $\gamma < J_2$, $E_{n, \pm}$ in Eq.\,(\ref{eq:NHeigenvalues}) are all real, and $\hat{H}_1$ resides in the $\mathcal{PT}$-symmetric unbroken phase (PTS). In contrast, for $\gamma > J_2$, a subset of $E_{n, \pm}$ emerge as complex-conjugate pairs, characterized by the $\mathcal{PT}$-symmetric broken phase (PTB). The quantum phase transition occurs at the critical point $\gamma \equiv J_2$, where the system spontaneously breaks $\mathcal{PT}$-symmetry. Within the PTB phase, a series of second-order EPs emerge at the condition 
$J_2\sqrt{n+1}=\gamma$, signifying simultaneous coalescence of eigenstates and eigenvalues in the subspace 
$\{\ket{\varphi_1(n)}, \ket{\varphi_2(n)}\}$.

Further, when the system is initialized in the subspace governed by $\hat{H}_1$, one could observe the $\mathcal{PT}$ phase transition via its dynamical behaviors. In the PTS phase, the  boson number dynamics exhibit oscillatory behavior, arising from the superposition of oscillating modes in multiple subspaces $\{\ket{\varphi_1(n)}, \ket{\varphi_2(n)}\}$ with distinct eigenfrequencies [Fig.\,\ref{fig4}(c)]. In the PTB phase, the system undergoes a transient period of mode competition dynamics, and then rapidly stabilizes at the largest-gain mode $\ket{\tilde{\psi}_{0,+}^R}$ [Fig.\,\ref{fig4}(e)].
We also compare the spin-boson entanglement dynamics, quantified by the von Neumann entropy of the reduced bosonic state $S(\hat{\rho}_\mathrm{boson})$, across distinct parameter regimes. As shown in [Fig.\,\ref{fig4}(d, f)], its oscillating-to-steady transition provides another signature of the $\mathcal{PT}$ phase transition.

\textit{Experimental proposal in the trapped-ion setup.}$-$Here, we propose an implementation protocol based on a trapped-ion platform. For a single trapped ion, it has internal states and external vibrations. The laser-ion interactions provide a mechanism to couple these two degrees of freedom.  When the interaction only involves two internal levels (ground state $\ket{g}$ and excited state $\ket{e}$) and the vibration amplitude is small, the system can be effectively modeled as a two-level spin interacting with a vibrational harmonic oscillator. Therefore, it forms a spin-boson system, where the spin states are defined as $\ket{\uparrow}=\ket{e}$ and $\ket{\downarrow}=\ket{g}$, and the vibrational Fock state $\ket{n}$ represents a state with $n$ vibrational quanta. 
	
To realize the semi-infinite SSH model described by the Hamiltonian in Eq.~(\ref{eq:one}), it is crucial to implement the resonant spin driving and the JC model coupling between the internal states and the vibrational mode. These two fundamental operations are widely used in the quantum state manipulation of trapped ions. In the interaction picture, the laser-ion interaction Hamiltonian \cite{Sasura2002} is given by
\begin{equation}
	\hat{H}_\mathrm{I} = \frac{\hbar \varOmega}{2} \hat{\sigma}_+ e^{-i\phi} e^{-i(\omega_L - \omega_0)t} \exp\left[i\eta (\hat{a}^{\dagger} e^{i\omega t} + \hat{a} e^{-i\omega t})\right] + \mathrm{h.c.},
	\label{D1}
\end{equation}
where $\varOmega$ is the Rabi frequency related to the incident laser's electric field strength, $\hat{\sigma}_{+}=\ket{e}\bra{g}$ is the spin raising operator, and $\hat{a}$, $\hat{a}^{\dagger} $ are the annihilation and creation operators of the vibrational mode. The incident laser's (angular) frequency and phase are denoted by $\omega_L$ and $\phi$, respectively. The transition frequency between the ion's two internal levels is $\omega_0$, and the ion's vibrational frequency is $\omega$. The Lamb-Dicke parameter is $\eta = k_L \sqrt{\hbar / (2\omega M)}$, where $M$ is the ion's mass and $k_L$ is the incident laser's wave vector. 
	
In the Lamb-Dicke regime, i.e., $\eta^2(2n+1) \ll 1$, and under the rotating wave approximation, by setting the laser frequency to the carrier resonance ($\omega_L=\omega_0$), resonant spin driving is achieved, while tuning it to the first red sideband ($\omega_L=\omega_0-\omega$) enables JC spin-vibration coupling. Simultaneously applying these two laser beams with their phases set to zero yields the following Hamiltonian,
\begin{equation}
	\hat{H} = J_1 (\hat{\sigma}_+ + \hat{\sigma}_-) + J_2 (\hat{a}^{\dagger} \hat{\sigma}_- + \hat{a} \hat{\sigma}_+).
	\label{D2}
\end{equation}
Here, $J_1 = \hbar \varOmega_c / 2$ represents the intracell coupling strength, $J_2 = \eta \hbar \varOmega_r / 2$, and $J_2 \sqrt{n+1}$ represents the intercell coupling strength. So adjusting the Rabi frequency $\varOmega _{c(r)}$ is equivalent to adjusting $J_{1(2)}$.
	
To implement the non-Hermitian Hamiltonian $\hat{H}_{\mathrm{NH}}$ described in Eq.~(\ref{eq:NHHamiltonian}), loss of $-i\gamma$ is needed for the $\ket{\uparrow}$ state and gain of $i\gamma$ for the $\ket{\downarrow}$ state. However, achieving gain is experimentally challenging. A practical solution is to apply a dissipation of $-2i\gamma$ exclusively to the $\ket{\uparrow}$ state, which is equivalent to incorporating a constant term $-i\gamma\hat{I}_2$ into the Hamiltonian, where $\hat{I}_2$ is the $2\times2$ identity operator. This method merely introduces an overall offset in the eigenspectrum, which does not affect the renormalized dynamics. Therefore, this purely dissipative approach can effectively simulates a non-Hermitian system with both gain and loss. Additionally, dissipation of the $\ket{\uparrow}$ state can be realized by coupling the energy level $\ket{e}$ to a short-lived state, inducing effective spontaneous emission at a rate of $2\gamma$, which is commonly used in non-Hermitian quantum systems \cite{PhysRevA.103.L020201a,PhysRevLett.126.083604a,quinn2023a}. See Appendix~\ref{AppendixD} for more technical details of the experimental scheme.

\textit{Conclusion.}$-$In conclusion, this work provides a rigorous characterization of Hermitian and non-Hermitian topological properties in semi-infinite anisotropic SSH models. We demonstrate the phenomena absent in isotropic systems---a more stable topological zero modes, non-Hermitian bound effect (NHBE), and a $\mathcal{PT}$ quantum phase transition within the restricted Hilbert space. Crucially, unlike prior studies confined to finite-dimensional subspaces with conserved excitations, our model inherently satisfies the thermodynamic limit while remaining analytically solvable. These results provide a methodological framework to investigate exotic phenomena related to diverse semi-infinite anisotropic topological systems based on FSL. Furthermore, all the predicted effects here, including NHBE and $\mathcal{PT}$ transitions, are directly observable on state-of-the-art quantum platforms.

\textit{Acknowledgments.}$-$We would like to acknowledge valuable discussions with
Yu-Yu Zhang. This work is supported by the National Natural Science Foundation of China (Grants No. 12174447, No. 12174448, No. 11574398), and the Innovation Program for Quantum Science and Technology (Grant No. 2021ZD0301605), the Science and Technology Innovation Program of Hunan under Grant No. 2022RC1194, and the Natural Science Foundation of Hunan Province under Grant No. 2023JJ10052.

\nocite{*}

\clearpage
\newpage
\onecolumngrid
\setcounter{page}{1}
\setcounter{equation}{0}
\renewcommand{\thefigure}{A\arabic{figure}}
\setcounter{figure}{0}
\appendix

\section*{Appendix}
\label{Appendix}
	
\section{Analytical solutions in the Hermitian regime}
\label{AppendixA}
Our anisotropic SSH chain is a general model based on the spin-boson system and can be solved analytically. For the Hamiltonian $\hat{H}$ of Eq.~(\ref{eq:one}) in the main text, we can write it in the following form, 
\begin{equation}
	\hat{H} = \hat{H}_{\mathrm{re}} + \hat{H}_{JC}, 
\end{equation}
where $\hat{H}_{\mathrm{re}} =J_1 \hat{\sigma}_x$ represents a resonant spin driving, and $\hat{H}_{JC}=J_2 (\hat{a} \hat{\sigma}_+ + \hat{a}^\dagger \hat{\sigma}_-)$ is the Hamiltonian of the solvable Jaynes-Cummings (JC) model \cite{Jaynes1963,Shore1993}. Applying the following unitary transformation to $\hat{H}$ with the displacement operator $\hat{D}(\alpha)=e^{\alpha \hat{a}^\dagger-\alpha^*\hat{a}}$, we find
\begin{equation}
	\begin{aligned}
		\hat{D}^{\dag}\left( \alpha \right) \hat{H}\hat{D}\left( \alpha \right) &=J_1 \hat{\sigma}_x + J_2\hat{D}^{\dag}\left( \alpha \right) \left( \hat{a}\hat{\sigma}_++\hat{a}^{\dag}\hat{\sigma}_- \right) \hat{D}\left( \alpha \right) \\
		&=J_1 \hat{\sigma}_x + J_2\left[ \left( \hat{a}+\alpha \right) \hat{\sigma}_++\left( \hat{a}^{\dag}+\alpha ^* \right) \hat{\sigma}_- \right] \\
		&=J_1 \hat{\sigma}_x + J_2\left( \alpha \hat{\sigma}_++\alpha ^*\hat{\sigma}_- \right) +J_2\left( \hat{a}\hat{\sigma}_++\hat{a}^{\dag}\hat{\sigma}_- \right),
	\end{aligned}	
\end{equation}
where $[\hat{a}, \hat{D}(\alpha)] = \alpha \hat{D}(\alpha)$ and $[\hat{a}^{\dagger}, \hat{D}(\alpha)] = \alpha^* \hat{D}(\alpha)$ are used in the derivation. Letting $\alpha = \alpha^*$, it can be simplified as
\begin{equation}
	\begin{split}
		\hat{D}^\dagger(\alpha) \hat{H} \hat{D}(\alpha) &= J_1 \hat{\sigma}_x + J_2 \alpha (\hat{\sigma}_+ + \hat{\sigma}_-) + J_2 (\hat{a} \hat{\sigma}_+ + \hat{a}^\dagger \hat{\sigma}_-) \\
		&= (J_1 + J_2 \alpha) \hat{\sigma}_x + J_2 (\hat{a} \hat{\sigma}_+ + \hat{a}^\dagger \hat{\sigma}_-).
	\end{split}
\end{equation}
To further simplify, we could set $J_1 + J_2 \alpha = 0$, and obtain
\begin{equation}
	\hat{D}^\dagger(\alpha) \hat{H} \hat{D}(\alpha) = J_2 (\hat{a} \hat{\sigma}_+ + \hat{a}^\dagger \hat{\sigma}_-) = \hat{H}_{JC}.
	\label{eq4}
\end{equation}
Therefore, $\hat{H}$ and $\hat{H}_{JC}$ are connected by a displacement transformation with $\alpha = \alpha^* = -J_1/J_2$. For the well-known $\hat{H}_{JC}$, its eigenvalue equations are 
\begin{equation}
	\begin{aligned}	
		\hat{H}_{JC} \ket{\phi_0} &= 0 \ket{\phi_0}, \\
		\hat{H}_{JC} \ket{\phi_{n, \pm}} &= \pm J_2 \sqrt{n+1} \ket{\phi_{n, \pm}},
		\label{JCeigenequation}
	\end{aligned}
\end{equation} 
where $\ket{\phi_0} = \ket{0, \downarrow}$ and $\ket{\phi_{n, \pm}} = (\ket{n, \uparrow} \pm \ket{n+1, \downarrow})/\sqrt{2} \quad(n = 0, 1, 2, \cdots, \infty)$. Multiplying Eq.~(\ref{JCeigenequation}) on the left by $\hat{D}(\alpha)$ and using Eq.~(\ref{eq4}), we have 
\begin{equation}
	\begin{aligned}	
		\hat{H} \ket{\varphi_0} &= 0 \ket{\varphi_0}, \\
		\hat{H} \ket{\varphi_{n, \pm}} &= \pm J_2 \sqrt{n+1} \ket{\varphi_{n, \pm}}.
		\label{JCeigeneq}
	\end{aligned}
\end{equation} 
Then, we obtain the eigenenergies of $\hat{H}$ as
\begin{equation}
	E = 0, \pm J_2 \sqrt{n+1}\quad(n = 0, 1, 2, \cdots, \infty)
\end{equation}
and the corresponding eigenstates as
\begin{equation}
	\begin{aligned}
		\ket{\varphi_0} &= \hat{D}(\alpha) \ket{\phi_0} = \hat{D}(\alpha) \ket{0, \downarrow} = \ket{\alpha, \downarrow}, \\
		\ket{\varphi_{n, \pm}} &= \hat{D}(\alpha) \ket{\phi_{n, \pm}} = \frac{1}{\sqrt{2}}\hat{D}(\alpha) (\ket{n, \uparrow}\pm \ket{n+1, \downarrow}),
	\end{aligned}
\end{equation}
which differ from the eigenstates of $\hat{H}_{JC}$ only by a displacement transformation $\hat{D}(\alpha)$. 

The structure of the eigenspectrum can be understood from the chiral symmetry. Since $\hat{H}$ satisfies the chiral symmetry $\mathcal{C} \hat{H} \mathcal{C}^{\dagger} = -\hat{H}$ ($\mathcal{C} = \hat{\sigma}_z$), we have $\{\mathcal{C}, \hat{H}\} = 0$, where $\{\cdot,\cdot\}$ represents the anticommutator. Combining wtih the eigenvalue equation
\begin{equation}
	\hat{H}\ket{\psi} =E\ket{\psi},
\end{equation}
we obtain
\begin{equation}
	\hat{H}\mathcal{C}\ket{\psi} =-\mathcal{C}\hat{H}\ket{\psi } =-\mathcal{C}E\ket{\psi} =-E\mathcal{C}\ket{\psi}. 
\end{equation}
Here, $\ket{\psi}$ and $ \mathcal{C}\ket{\psi}$ are chiral partners with opposite eigenenergies ($E$ and $-E$). The action of $\mathcal{C}$ on $\ket{\varphi_0}$ and $\ket{ \varphi_{n,\pm}}$ yields:
\begin{equation}
	\begin{aligned}
		\mathcal{C} \ket{\varphi_{n, \pm}} &= \ket{\varphi_{n, \mp}}, \\
		\mathcal{C} \ket{\alpha, \downarrow} &= -\ket{\alpha, \downarrow}. 
	\end{aligned}
\end{equation}
Thus, $\ket{\varphi_{n, +}}$ and $\ket{\varphi_{n, -}}$ are chiral partners, and their corresponding eigenenergies are symmetric about 0. However, for the topological zero mode $\ket{\alpha, \downarrow}$, its chiral partner $-\ket{\alpha, \downarrow}$ only differs by a global $\pi$ phase, so its chiral partner is itself. Therefore, our anisotropic SSH chain has only one zero mode, in consistent with the fact that our model is semi-infinite and has only one left boundary.

\section{$\mathcal{PT}$ Symmetry and $\mathcal{PT}$ Phase Transition in the Subspace} 
\label{AppendixB}
Motivated by the analytical solution in the Hermitian realm, and utilizing the displaced Fock-state basis, we find that the non-Hermitian Hamiltonian $\hat{H}_\mathrm{NH}$ of Eq.~(\ref{eq:NHHamiltonian}) in the main text satisfies the following relations:
\begin{equation}
	\begin{aligned}
		\hat{H}_\mathrm{NH} \ket{\alpha, \downarrow} &= i\gamma \ket{\alpha, \downarrow}, \\
		\hat{H}_\mathrm{NH} \begin{bmatrix} \ket{\varphi_1(n)} \\ \ket{\varphi_2(n)} \end{bmatrix} &= \begin{bmatrix} -i\gamma & J_2 \sqrt{n+1} \\ J_2 \sqrt{n+1} & i\gamma \end{bmatrix} \begin{bmatrix} \ket{\varphi_1(n)} \\ \ket{\varphi_2(n)} \end{bmatrix}.
	\end{aligned}
\end{equation}
Here, $\ket{\varphi_1(n)} = \hat{D}(\alpha) \ket{n, \uparrow}$, $\ket{\varphi_2(n)} = \hat{D}(\alpha) \ket{n+1, \downarrow}$. Using $\ket{\varphi_0}$, $\ket{\varphi_1(n)}$ and $\ket{\varphi_2(n)}$ ($ n=0,1,2,\cdots,\infty$) as basis vectors,
$\hat{H}_\mathrm{NH}$ can be written in the following form:
\begin{equation}
	\hat{H}_\mathrm{NH} = \hat{H}_0 \oplus \hat{H}_1,
	\label{eq13}
\end{equation}
where $\hat{H}_0=i\gamma\ket{\varphi_0}\bra{\varphi_0}$ and 
\begin{equation}
	\hat{H}_1=\bigoplus_{n=0}^{\infty}\left[J_2\sqrt{n+1}\hat{X}_n-i\gamma\hat{Z}_n\right]. 
\end{equation}
Here $\hat{X}_n=\ket{\varphi_1(n)}\bra{\varphi_2(n)}+\ket{\varphi_2(n)}\bra{\varphi_1(n)}$ and $\hat{Z}_n=\ket{\varphi_1(n)}\bra{\varphi_1(n)}-\ket{\varphi_2(n)}\bra{\varphi_2(n)}$ represent Pauli $x$- and $z$-operators in the subspace $\{\ket{\varphi_1(n)},\ket{\varphi_2(n)}\}$. Defining
\begin{equation}
	\begin{aligned}
		\boldsymbol{a} &= (\ket{\varphi_0}, \ket{\varphi_1(0)}, \ket{\varphi_2(0)}, \ket{\varphi_1(1)}, \ket{\varphi_2(1)}, \cdots), \\
		\boldsymbol{b} &= (\bra{\varphi_0}, \bra{\varphi_1(0)}, \bra{\varphi_2(0)}, \bra{\varphi_1(1)}, \bra{\varphi_2(1)}, \cdots)^T,
	\end{aligned}
\end{equation}
Eq.~(\ref{eq13}) can be written as
\begin{equation}
	\hat{H}_\mathrm{NH} = \boldsymbol{a} \mathbb{H} \boldsymbol{b},
\end{equation}
where $\mathbb{H} = \mathbb{H}_0 \oplus \mathbb{H}_1$, $\mathbb{H}_0 = i\gamma$, $\mathbb{H}_1 = \bigoplus_{n=0}^{\infty} h_n$, and
\begin{equation}
	h_n = \begin{bmatrix} -i\gamma & J_2 \sqrt{n+1} \\ J_2 \sqrt{n+1} & i\gamma \end{bmatrix}.
\end{equation}
Thus, $\hat{H}_\mathrm{NH}$ becomes a direct sum of a scalar and infinitely many two-dimensional Hamiltonians decoupled with each other. The eigenvalue of the scalar space $\mathbb{H}_0$ is $i\gamma$. And after diagonalizing $h_n$, we obtain that the eigenvalues of the subspace $h_n$ are $\pm \sqrt{J_2^2(n+1) - \gamma^2}$, with the eigenvectors
\begin{equation}
	\ket{\pm}_n = \begin{bmatrix} J_2 \sqrt{n+1} \\ i\gamma \pm \sqrt{J_2^2(n+1) - \gamma^2} \end{bmatrix}.
\end{equation}
Therefore, the eigenenergies of $\hat{H}_\mathrm{NH}$ are $E_0 = i\gamma$, $E_{n, \pm} = \pm \sqrt{J_2^2(n+1) - \gamma^2}$, and the corresponding eigenstates are
\begin{equation}
	\begin{aligned}
		\ket{\varphi_0} &= \ket{\alpha, \downarrow}, \\
		\ket{\psi_{n, \pm}} &= J_2 \sqrt{n+1} \ket{\varphi_1(n)} + \left[i\gamma \pm \sqrt{J_2^2(n+1) - \gamma^2}\right] \ket{\varphi_2(n)},\quad n = 0, 1, \cdots, \infty.
	\end{aligned}
	\label{eigenstates}
\end{equation}
When $\gamma = 0$, it corresponds exactly to the solution in the Hermitian case. For simplicity, the eigenstates $|\psi_{n, \pm}\rangle$ are not normalized here. Because the right eigenstates of a non-Hermitian Hamiltonian are generally not orthogonal, one typically utilizes the biorthogonal basis formed by the right and left eigenstates, which satisfy the biorthogonal normalization condition (see Appendix~\ref{AppendixC}).

\begin{figure}[b]
	\includegraphics{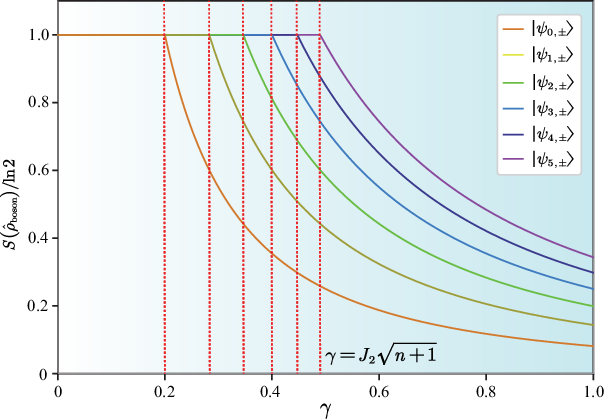}% Here is how to import EPS art
	\caption{\label{A1} The Von Neumann entropy $S(\hat{\rho}_{\mathrm{boson}})/\ln 2$ of the eigenstates $\ket{\psi_{n,\pm}}$ in Eq.~(\ref{eigenstates}) as a function of $\gamma$. For simplicity, $n$ ranges from 0 to 5, with each value corresponding to a distinct color. The parameters are set to $J_1 = 1$ and $J_2 = 0.2$. When $\gamma \leq J_2 \sqrt{n+1}$, $\ket{\psi_{n,\pm}}$ are maximally entangled, with $S(\hat{\rho}_{\mathrm{boson}})/\ln 2 = 1$. When $\gamma > J_2 \sqrt{n+1}$, $S(\hat{\rho}_{\mathrm{boson}})/\ln 2$ decreases monotonically with $\gamma/J_2$. The red dashed lines indicate EP ($\gamma = J_2 \sqrt{n+1}$) of each subspace $h_n$ in $\hat{H}_1$. As $n$ increases, the distance between them decreases.
	}
\end{figure}

For a non-Hermitian system satisfying parity-time ($\mathcal{PT}$) symmetry, its eigenenergies can be real numbers \cite{Bender1998,Bender1999,Mostafazadeh2002}. In quantum mechanics, $\mathcal{P}: \hat{x} \rightarrow -\hat{x}, \hat{p} \rightarrow -\hat{p}$ is the parity operator acting like spatial reflection. $\mathcal{T}:\hat{p} \rightarrow -\hat{p}, i \rightarrow -i$ is the time-reversal operator. Recently, it has been proved that the definitions of $\mathcal{P}$ and $\mathcal{T}$ operators can be generalized in the following way: in a finite-dimensional Hilbert space, if a linear operator $\mathcal{P}$ and an antilinear operator $\mathcal{T}$ satisfy $\mathcal{P}^2 = \hat{I}$, $\mathcal{T}^2 = \hat{I}$ and $[\mathcal{P},\mathcal{T}] = 0$ simultaneously, then $\mathcal{P}$ and $\mathcal{T}$ can be defined as the generalized parity operator and time-reversal operator, respectively \cite{huang2018,huang2018b}. As mentioned above, the Hilbert space of $\hat{H}_{\mathrm{NH}}$ can be decomposed into a direct sum of finite-dimensional subspaces. We find that in each of these subspaces $\{\ket{\varphi_1(n)}, \ket{\varphi_2(n)}\}$, the generalized definition of the $\mathcal{PT}$ operator applies. Consequently, even though $\hat{H}_{\mathrm{NH}}$ is infinite-dimensional, the generalized $\mathcal{PT}$ definition remains valid in the subspace orthgonal to the bound state $\ket{\alpha,\downarrow}$. Therefore, we can define $\mathcal{P}= \bigoplus_{n=0}^{\infty} \hat{X}_n$ and $\mathcal{T}=\mathcal{K}$, where $\mathcal{K}$ is the complex conjugation operator. Then, we obtain the $\mathcal{PT}$ operator as
\begin{equation}
	\mathcal{PT} = \bigoplus_{n=0}^{\infty} \hat{X}_n \mathcal{K}. 
\end{equation}
It can be easily checked that
\begin{equation}
	\begin{aligned}	
		(\mathcal{PT}) \hat{H}_1 (\mathcal{PT})^{-1} &= \bigoplus_{n=0}^{\infty}{\left( \ket{\varphi_1(n)} ,\ket{\varphi_2(n)} \right) \left[ \begin{matrix}
				0&		1\\
				1&		0\\
			\end{matrix} \right]}h_{n}^{*}\left[ \begin{matrix}
			0&		1\\
			1&		0\\
		\end{matrix} \right] \left( \begin{array}{c}
			\bra{\varphi_1(n)}\\
			\bra{\varphi_2(n)}\\
		\end{array} \right) \\
		&= \bigoplus_{n=0}^{\infty}{\left( \ket{\varphi_1(n)},\ket{\varphi_2(n)} \right) h_n\left( \begin{array}{c}
				\bra{\varphi_1(n)}\\
				\bra{\varphi_2(n)}\\
			\end{array} \right)} \\
		&= \hat{H}_1,
	\end{aligned}	  
\end{equation}
which is equivalent to $[\hat{H}_1, \mathcal{PT}] = 0$. Combining with the eigenvalue equation $\hat{H}_1 \ket{\psi_n} = E_n \ket{\psi_n}$, we obtain
\begin{equation}
	\hat{H}_1\mathcal{P}\mathcal{T}\ket{\psi_n}=\mathcal{P}\mathcal{T}\hat{H}_1\ket{\psi_n}=\mathcal{P}\mathcal{T}E_n\ket{\psi_n}=\mathcal{P}\mathcal{T}E_n\left( \mathcal{P}\mathcal{T} \right) ^{-1}\mathcal{P}\mathcal{T}\ket{\psi_n}=E_{n}^{*}\mathcal{P}\mathcal{T}\ket{\psi_n}.
\end{equation}
It implies that for an eigenenergy $E_n$ with corresponding eigenstate $\ket{\psi_n}$, there exists an eigenenergy $E_n^*$ with corresponding eigenstate $\mathcal{PT} \ket{\psi_n}$. If the system's eigenstates are $\mathcal{PT}$ symmetric, such that $\mathcal{PT} \ket{\psi_n} = \lambda_n^*\ket{\psi_n}$, where the modulus of the eigenvalue satisfies $\left|\lambda_n\right|=1$, then $E_n = E_n^*$, so all eigenenergies are real, defining $\mathcal{PT}$-symmetric phase. Otherwise, it is defined as $\mathcal{PT}$-symmetric broken phase.

As discussed above, $\hat{H}_1$ satisfies $\mathcal{PT}$ symmetry. This can also be deduced directly from the eigenenergies. The eigenenergies of $\hat{H}_\mathrm{NH}$ are $E_0 = i\gamma$ and $E_{n, \pm} = \pm \sqrt{J_2^2(n+1) - \gamma^2}$. Apart from the isolated $i\gamma$, the eigenenergies $E_{n, \pm}$ of the subspace $\hat{H}_1$ are symmetric about both the real and imaginary axes in the complex plane. Whether $E_{n, \pm}$ are real or purely imaginary depends on the ratio $\gamma/J_2$. Specifically, when $\gamma \leq J_2$, all $E_{n, \pm}$ are real, placing $\hat{H}_1$ in the $\mathcal{PT}$-symmetric phase. When $\gamma > J_2$, $E_{n, \pm}$ become partially purely imaginary and partially real, and $\hat{H}_1$ is in the $\mathcal{PT}$-symmetry-broken phase. If $(\gamma/J_2)^2 \in \mathbb{N}_+$, where $\mathbb{N}_+$ denotes the set of positive integers, there exists an $n_c$ satisfying $J_2 \sqrt{n_c + 1} = \gamma$, corresponding to the two-dimensional subspace $h_{n_c}$ at the exceptional point (EP), where both the eigenvalues and eigenvectors in $h_{n_c}$ coalesce, i.e.,
\begin{equation}
	E_{n_c, +} = E_{n_c, -} = 0, \quad \ket{+}_{n_c} = \ket{-}_{n_c} = \begin{bmatrix} 1 \\ i \end{bmatrix}.
\end{equation}
Note that as $\gamma/J_2$ varies, each subspace $h_n$ of $\hat{H}_1$ exhibits an EP, while $\hat{H}_1$ has only one phase transition point at $\gamma/J_2 = 1$ [Fig.~\ref{fig4}(a) and~\ref{fig4}(b) in the main text].
Therefore, as $\gamma$ changes from $\gamma < J_2$ to $\gamma > J_2$, $\hat{H}_1$ undergoes a $\mathcal{PT}$ phase transition. Furthermore, we find that when $\gamma \le J_2 \sqrt{n+1}$,
\begin{equation}
	\dfrac{\left|J_2 \sqrt{n+1}\right|}{\left|i\gamma \pm \sqrt{J_2^2(n+1) - \gamma^2}\right|} =1,
\end{equation}
where $|\cdot|$ represents the modulus of a complex number. Consequently, the eigenstates $\ket{\psi_{n, \pm}}$ in Eq.~(\ref{eigenstates}) of $\hat{H}_1$ are equal superpositions of $\ket{\varphi_1(n)}$ and $\ket{\varphi_2(n)}$, representing maximally entangled states between the spin and bosonic mode. Their entanglement can be measured by the Von Neunmann entropy of the reduced bosonic state $S(\hat{\rho}_{\mathrm{boson}})$, where $S(\hat{\rho})= -Tr(\hat{\rho}\mathrm{ln}\hat{\rho})$. The entanglement of $\ket{\psi_{n,\pm}}$ decreases monotonically with $\gamma/J_2$ when $\gamma > J_2 \sqrt{n+1}$ (Fig.~\ref{A1}).

\section{The Non-Hermitian Dynamics} \label{AppendixC}
For a general non-Hermitian Hamiltonian $\hat{\mathcal{H}}\ne \hat{\mathcal{H}}^{\dag}$, the eigenvalue equations of $\hat{\mathcal{H}}$ and $\hat{\mathcal{H}}^{\dag}$ are given by \cite{Ashida2020,Brody2013}
\begin{equation}
	\hat{\mathcal{H}}\ket{u _{n}^{R}} =\epsilon_n\ket{u _{n}^{R}} ,\ \hat{\mathcal{H}}^{\dag}\ket{u _{n}^{L}} =\epsilon_{n}^{*}\ket{u _{n}^{L}},
\end{equation}
where $\epsilon_n$ is the $n$th eigenenergy, and $\ket{u _{n}^{R}}$ and $\ket{u _{n}^{L}}$ are the
right and left eigenstates that satisfy the completeness
relation $\sum_n{\ket{u_{n}^{R}}\bra{u_{n}^{L}} =\hat{I}}$ and the biorthonormal relation $\braket{u_{m}^{L}}{u_{n}^{R}} =\delta _{mn}$. Here, eigenstates are no longer normalized, i.e., $\braket{u_{n}^{R}}{u_{n}^{R}}\ne 1$ and $\braket{u_{n}^{L}}{u_{n}^{L}}\ne 1$.
In Appendix~\ref{AppendixB}, we have analytically derived the right eigenstates of $\hat{H}_\mathrm{NH}$. To address the non-Hermitian dynamics, we need employ the biorthogonal basis and the first task is to derive the left eigenstates. For $\hat{H}_\mathrm{NH}^{\dagger}$, we have
\begin{equation}
	\begin{aligned}
		\hat{H}_\mathrm{NH}^{\dagger} \ket{\alpha, \downarrow} &= -i\gamma \ket{\alpha, \downarrow}, \\
		\hat{H}_\mathrm{NH}^{\dagger} \begin{bmatrix} \ket{\varphi_1(n)} \\ \ket{\varphi_2(n)} \end{bmatrix} &= \begin{bmatrix} i\gamma & J_2 \sqrt{n+1} \\ J_2 \sqrt{n+1} & -i\gamma \end{bmatrix} \begin{bmatrix} \ket{\varphi_1(n)} \\ \ket{\varphi_2(n)} \end{bmatrix}.
	\end{aligned}
\end{equation}
Therefore, the eigenstates of $\hat{H}^{\dagger}_\mathrm{NH}$ are
\begin{equation}
	\begin{aligned}
		\ket{\varphi_0^L} &= \ket{\alpha, \downarrow}, \\
		\ket{\psi_{n, \pm}^L} &= J_2 \sqrt{n+1} \ket{\varphi_1(n)} + (i\gamma + E_{n, \pm})^* \ket{\varphi_2(n)},\quad n = 0, 1,\cdots,\infty.
	\end{aligned}
\end{equation}
Then, the biorthogonal normalized right eigenstates are given by
\begin{equation}
	\ket{\tilde{\varphi}_0^R} = \ket{\alpha, \downarrow}, \quad \ket{\tilde{\psi}_{n, \pm}^R} = \frac{\ket{\psi_{n, \pm}^R}}{\sqrt{\braket{\psi_{n, \pm}^L}{\psi_{n, \pm}^R}}},
\end{equation}
and the biorthogonal normalized left eigenstates are
\begin{equation}
	\ket{\tilde{\varphi}_0^L} = \ket{\alpha, \downarrow}, \quad \ket{\tilde{\psi}_{n, \pm}^L} = \frac{\ket{\psi_{n, \pm}^L}}{\sqrt{\braket{\psi_{n, \pm}^R}{\psi_{n, \pm}^L}}}.
\end{equation}
Consequently, we have derived the biorthogonal basis for $\hat{H}_\mathrm{NH}$, satisfying
\begin{equation}
	\braket{\tilde{\varphi}_0^L}{\tilde{\varphi}_0^R} = 1, \quad \braket{\tilde{\varphi}_0^L}{\tilde{\psi}_{n, g}^R} = \braket{\tilde{\psi}_{n, g}^L}{\tilde{\varphi}_0^R} = 0, \quad \braket{\tilde{\psi}_{n, g}^L}{\tilde{\psi}_{n^{\prime}, g^{\prime}}^R} = \delta_{n n^{\prime}}\delta_{g g^{\prime}}\quad(g,g^{\prime}=\pm).
\end{equation}
When $(\gamma/J_2)^2 \notin \mathbb{N}_+$, the completeness relation for $\hat{H}_\mathrm{NH}$ is
\begin{equation}
	\ket{\alpha ,\downarrow} \bra{\alpha ,\downarrow}+\sum_{n=0}^{\infty}{\sum_{g=+,-}{\ket{\tilde{\psi}_{n,g}^{R}} \bra{\tilde{\psi}_{n,g}^{L}}}=\hat{I}},
\end{equation}
where $\hat{I}$ is the identity operator in the infinite-dimensional Hilbert space. When $(\gamma/J_2)^2 \in \mathbb{N}_+$, a special situation arises, as the corresponding eigenstates of $h_{n_c}$ coalesce, i.e.,
\begin{equation}
	\ket{\psi_{n_c, +}} = \ket{\psi_{n_c, -}} = \ket{\varphi_1(n_c)} + i\ket{\varphi_2(n_c)}.
\end{equation}
This eigenstate does not satisfy the completeness relation within the subspace $h_{n_c}$. However, since $\ket{\varphi_1(n_c)}$ and $\ket{\varphi_2(n_c)}$ form an orthonormal basis for $h_{n_c}$, they naturally satisfy the completeness relation,
\begin{equation}
	\ket{\varphi_1(n_c)} \bra{\varphi_1(n_c)} + \ket{\varphi_2(n_c)} \bra{\varphi_2(n_c)} = 0 \oplus \bigoplus_{n=0}^{n_c-1} \mathbb{O} \oplus \hat{I}_2 \oplus \bigoplus_{n=n_c+1}^{\infty} \mathbb{O},
\end{equation}
where $\mathbb{O}$ is the zero matrix for the subspace $h_n$, and $\hat{I}_2$ is the identity matrix for $h_{n_c}$. Therefore, we can express the completeness relation as
\begin{equation}
	\ket{\alpha ,\downarrow} \bra{\alpha ,\downarrow }+\sum_{n=0,\ne n_c}^{\infty}{\sum_{g=+,-}{\ket{\tilde{\psi}_{n,g}^{R}} \bra{\tilde{\psi}_{n,g}^{L}}}}+\ket{\varphi _1(n_c)} \bra{\varphi _1(n_c)}+\ket{\varphi _2(n_c)} \bra{\varphi _2(n_c)}=\hat{I}.
\end{equation}

From Appendix~\ref{AppendixB}, we have $\hat{H}_\mathrm{NH} = \boldsymbol{a} \mathbb{H} \boldsymbol{b}$, so the time evolution operator (setting $\hbar = 1$) is
\begin{equation}
	\hat{U}(t) = e^{-i\hat{H}_\mathrm{NH}t} = \boldsymbol{a} e^{-i\mathbb{H}t} \boldsymbol{b},
\end{equation}
where $e^{-i\mathbb{H}t} = e^{\gamma t} \oplus \bigoplus_{n=0}^{\infty} e^{-i h_n t}$. For a given initial state $|\varPsi(0)\rangle$, we can express it in terms of the eigenstates using the biorthogonal basis [when $(\gamma/J_2)^2 \notin \mathbb{N}_+$],
\begin{equation}
	\ket{\varPsi(0)} =c_0\ket{\alpha ,\downarrow} +\sum_{n=0}^{\infty}{\sum_{g=+,-}{c_{n,g}\ket{\tilde{\psi}_{n,g}^{R}}}},
\end{equation}
where the superposition coefficients are given by
\begin{equation}
	c_0 = \braket{\alpha, \downarrow}{\varPsi(0)}, \quad c_{n, +} = \braket{\tilde{\psi}_{n, +}^L}{\varPsi(0)}, \quad c_{n, -} = \braket{\tilde{\psi}_{n, -}^L}{\varPsi(0)}.
\end{equation}
Therefore, the time-evolved wavefunction at time $t$ is
\begin{equation}
	\begin{aligned}
		\ket{\varPsi(t)} &=\hat{U}(t) \ket{\varPsi(0)} =e^{-i\hat{H}_\mathrm{NH}t}\left[ c_0\ket{\alpha ,\downarrow} +\sum_{n=0}^{\infty}{\sum_{g=+,-}{c_{n,g}\ket{\tilde{\psi}_{n,g}^{R}}}} \right] \\
		&=\boldsymbol{a}e^{-i\mathbb{H}t}\boldsymbol{b}\left[ c_0\boldsymbol{a}\left( 1,0,\cdots ,0 \right) ^T+\sum_{n=0}^{\infty}{\sum_{g=+,-}{c_{n,g}}}\boldsymbol{a}\left( 0,\cdots ,J_2\sqrt{n+1},i\gamma +E_{n,g},\cdots ,0 \right) ^T/\sqrt{\braket{\psi _{n,g}^{L}}{\psi _{n,g}^{R}}} \right] \\
		&=c_0\boldsymbol{a}e^{-i\mathbb{H}t}\left( 1,0,\cdots ,0 \right) ^T+\sum_{n=0}^{\infty}{\sum_{g=+,-}{c_{n,g}}}\boldsymbol{a}e^{-i\mathbb{H}t}\left( 0,\cdots ,J_2\sqrt{n+1},i\gamma +E_{n,g},\cdots ,0 \right) ^T/\sqrt{\braket{\psi _{n,g}^{L}}{\psi _{n,g}^{R}}} \\
		&=c_0e^{\gamma t}\boldsymbol{a}\left( 1,0,\cdots ,0 \right) ^T+\sum_{n=0}^{\infty}{\sum_{g=+,-}{c_{n,g}}}e^{-iE_{n,g}t}\boldsymbol{a}\left( 0,\cdots ,J_2\sqrt{n+1},i\gamma +E_{n,g},\cdots ,0 \right) ^T/\sqrt{\braket{\psi _{n,g}^{L}}{\psi _{n,g}^{R}}} \\
		&=c_0e^{\gamma t}\ket{\alpha ,\downarrow} +\sum_{n=0}^{\infty}{\sum_{g=+,-}{c_{n,g}}}e^{-iE_{n,g}t}\ket{\tilde{\psi}_{n,g}^{R}}.
	\end{aligned}
\end{equation}
In the derivation, we transform the biorthogonal basis to the $\boldsymbol{a}, \boldsymbol{b}$ basis, perform the analysis, and subsequently revert to the biorthogonal basis. When $(\gamma/J_2)^2 \in \mathbb{N}_+$, the initial state is given by
\begin{equation}
	\ket{\varPsi(0)} =c_0\ket{\alpha ,\downarrow} +\sum_{n=0,\ne n_c}^{\infty}{\sum_{g=+,-}{c_{n,g}\ket{ \tilde{\psi}_{n,g}^{R}}}}+c_1\ket{\varphi _1(n_c)} +c_2\ket{\varphi _2(n_c)},
\end{equation}
where $c_1 = \braket{\varphi_1(n_c)}{\varPsi(0)}$ and $c_2 = \braket{\varphi_2(n_c)}{\varPsi(0)}$. Because the 2D subspace
\begin{equation}
	h_{n_c} = \gamma \begin{bmatrix} -i & 1 \\ 1 & i \end{bmatrix}
\end{equation}
is at the EP, we can apply a Jordan normalization to $h_{n_c}$: $h_{n_c} = PJP^{-1}$, where
\begin{equation}
	P = \begin{bmatrix} 1 & 0 \\ i & 1 \end{bmatrix}, \quad P^{-1} = \begin{bmatrix} 1 & 0 \\ -i & 1 \end{bmatrix}, \quad J = \gamma \begin{bmatrix} 0 & 1 \\ 0 & 0 \end{bmatrix}.
\end{equation}
The corresponding time evolution operator for $h_{n_c}$ is $\left( \ket{\varphi _1(n_c)} ,\ket{\varphi _2(n_c)} \right) e^{-ih_{n_c}t}\left( \bra{\varphi _1(n_c)},\bra{\varphi _2(n_c)} \right) ^T,$ with
\begin{equation}
	e^{-i h_{n_c} t} = e^{-i PJP^{-1} t} = P e^{-i J t} P^{-1} = P (1 - i t J) P^{-1} = \begin{bmatrix} 1 - \gamma t & -i\gamma t \\ -i\gamma t & 1 + \gamma t \end{bmatrix}.
\end{equation}
Therefore, the time-evolved wavefunction is
\begin{equation}
	\begin{aligned}
		\ket{\varPsi(t)} &=\hat{U}(t) \ket{\varPsi(0)}\\ 
		&=c_0e^{\gamma t}\ket{\alpha ,\downarrow} +\sum_{n=0,\ne n_c}^{\infty}{\sum_{g=+,-}{c_{n,g}e^{-iE_{n,g}t}\ket{ \tilde{\psi}_{n,g}^{R}}}}\\
		&\quad+\left[ (1-\gamma t) c_1-i\gamma tc_2 \right] \ket{ \varphi _1(n_c)} +\left[ -i\gamma tc_1+(1+\gamma t) c_2 \right] \ket{\varphi _2\left( n_c \right)}.
	\end{aligned}
\end{equation}
Since the time evolution operator in the non-Hermitian case is not unitary [$\hat{U}(t) \hat{U}^{\dagger}(t) \neq \hat{I}$], $\ket{\varPsi(t)}$ no longer satisfies the normalization condition, even if the initial state is normalized. So it needs to be renormalized.

As mentioned in the main text, when renormalizing $\ket{\varPsi(t)}$, the specific basis must be considered. If the basis is orthonormal, such as the Fock basis, the renormalized state is given by $\ket{\varPsi^{\prime}(t)} = \ket{\varPsi(t)}/\sqrt{\braket{\varPsi(t)}}$. Since the Fock basis naturally satisfies the completeness relation,
\begin{equation}
	\sum_{n=0}^\infty \sum_{l=\uparrow,\downarrow} \ket{n,l}\bra{n,l} = \hat{I}, 
\end{equation} 
the normalized initial state $\ket{\varPsi(0)}$ can be expanded in the Fock basis as
\begin{equation}
	\ket{\varPsi(0)}=\sum_{n=0}^\infty \sum_{l=\uparrow,\downarrow} c_{n,l} \ket{n,l},
\end{equation}
where $c_{n,l} = \braket{n,l}{\varPsi(0)}$. Thus, it is straightforward to show that
\begin{equation}
	\begin{aligned}
		\sum_{n=0}^\infty \sum_{l=\uparrow,\downarrow} |c_{n,l}|^2 &= \sum_{n=0}^\infty \sum_{l=\uparrow,\downarrow} \braket{\varPsi(0)}{n,l}\braket{n,l}{\varPsi(0)} =\bra{\varPsi(0)}\sum_{n=0}^\infty \sum_{l=\uparrow,\downarrow} \ket{n,l}\bra{n,l}\ket{\varPsi(0)} \\
		&= \braket{\varPsi(0)} = 1.
	\end{aligned}
\end{equation}
The projection probability of $\ket{\varPsi^{\prime}(t)}$ onto $\ket{n,l}$ is defined as 
\begin{equation}
	P_{n,l}(t) = |c_{n,l}(t)|^2 = |\braket{n,l}{\varPsi^{\prime}(t)}|^2.
\end{equation} 
Since $\braket{\varPsi^{\prime}(t)}{\varPsi^{\prime}(t)}=1$, we have
\begin{equation} 
	\sum_{n=0}^\infty \sum_{l=\uparrow,\downarrow} P_{n,l}(t) = 1. 
\end{equation}
Thus, the boson number distribution at time $t$ is given by
\begin{equation}
	P_n(t) = \sum_{l=\uparrow,\downarrow} P_{n,l}(t).
\end{equation}

If calculating the projection probability of a normalized state $\ket{\varPsi^{\prime}(t)}$ onto the eigenmodes $\ket{\tilde{\psi}_{n,g}^R}$, the completeness relation of the biorthogonal basis needs to be used. It is readily obtained
\begin{equation}
	\begin{aligned}
		\braket{\varPsi^{\prime}(t)} &= \bra{\varPsi^{\prime}(t)}\left(\ket{\alpha, \downarrow}\bra{\alpha, \downarrow} + \sum_{n=0}^\infty \sum_{g=+,-} \ket{\tilde{\psi}_{n,g}^R}\bra{\tilde{\psi}_{n,g}^L}\right)\ket{\varPsi^{\prime}(t)} \\
		&= \braket{\varPsi^{\prime}(t)}{\alpha, \downarrow}\braket{\alpha, \downarrow}{\varPsi^{\prime}(t)} + \sum_{n=0}^\infty \sum_{g=+,-} \braket{\varPsi^{\prime}(t)}{\tilde{\psi}_{n,g}^R}\braket{\tilde{\psi}_{n,g}^L}{\varPsi^{\prime}(t)} \\
		&= 1.
	\end{aligned}
\end{equation}
However, if we consider $\braket{\varPsi^{\prime}(t)}{\tilde{\psi}_{n,g}^R}\braket{\tilde{\psi}_{n,g}^L}{\varPsi^{\prime}(t)}$ as the projection probability, this will lead to complex probabilities, which is inconsistent with actual measurements. In quantum theory, the norm of a state is closely related to probabilistic interpretations of measurement outcomes. This problem can be fixed by introducing the associated state $\ket{\bar{\varPsi}(t)}$ of $\ket{\varPsi(t)}$ \cite{Brody2013},
\begin{equation}
	\ket{\bar{\varPsi}(t)} = c_0 e^{\gamma t} \ket{\alpha, \downarrow} + \sum_{n=0}^\infty \sum_{g=+,-} c_{n,g} e^{-i E_{n,g} t} \ket{\tilde{\psi}_{n,g}^L}.
\end{equation}
So, the inner product of $\ket{\varPsi(t)}$ with itself is defined as $\braket{\bar{\varPsi}(t)}{\varPsi(t)}$, and the renormalized state is $\ket{\varPsi(t)}/\sqrt{\braket{\bar{\varPsi}(t)}{\varPsi(t)}}$, whose projection probability onto the eigenmode is given by
\begin{equation}
	\begin{aligned}
		P_0(t) &= \frac{\braket{\bar{\varPsi}(t)}{\alpha, \downarrow}\braket{\alpha, \downarrow}{\varPsi(t)}}{\braket{\bar{\varPsi}(t)}{\varPsi(t)}} = \frac{|\braket{\alpha, \downarrow}{\varPsi(t)}|^2}{\braket{\bar{\varPsi}(t)}{\varPsi(t)}} = \frac{|c_0|^2 e^{2\gamma t}}{|c_0|^2 e^{2\gamma t} + \sum_{n=0}^\infty \sum_{g=+,-} |c_{n,g}|^2 e^{-i(E_{n,g}-E_{n,g}^*)t}}, \\
		P_{n,g}(t) &= \frac{\braket{\bar{\varPsi}(t)}{\tilde{\psi}_{n,g}^R}\braket{\tilde{\psi}_{n,g}^L}{\varPsi(t)}}{\braket{\bar{\varPsi}(t)}{\varPsi(t)}} = \frac{\left|\braket{\tilde{\psi}_{n,g}^L}{\varPsi(t)}\right|^2}{\braket{\bar{\varPsi}(t)}{\varPsi(t)}} = \frac{|c_{n,g}|^2 e^{-i(E_{n,g}-E_{n,g}^*)t}}{|c_0|^2 e^{2\gamma t} + \sum_{n=0}^\infty \sum_{g=+,-} |c_{n,g}|^2 e^{-i(E_{n,g}-E_{n,g}^*)t}}.
	\end{aligned}	
\end{equation}
And it is readily shown that
\begin{equation}
	P_0(t)+\sum_{n=0}^\infty \sum_{g=+,-}P_{n,g}(t)=1.
\end{equation}	

The projection probability here actually reflects the relative weight of the eigenmode in $\ket{\varPsi(t)}$. The two methods for renormalizing $\ket{\varPsi(t)}$ are essentially aimed at finding the relative weights of the corresponding modes in $\ket{\varPsi(t)}$, differing only due to the different completeness relations of the basis vectors. For Figure.~\ref{fig3}(g) in the main text, when $c_0 \neq 0$, the time $\tau$ for $\ket{\varPsi(t)}$ to reach the bound state is defined such that $P_0(\tau) \to 1$ [the criterion for numerical calculation is $P_0(\tau) > 0.999999$].

\section{Technical Details of the Experimental Protocol}
\label{AppendixD}
	
\begin{figure}[b]
	\includegraphics{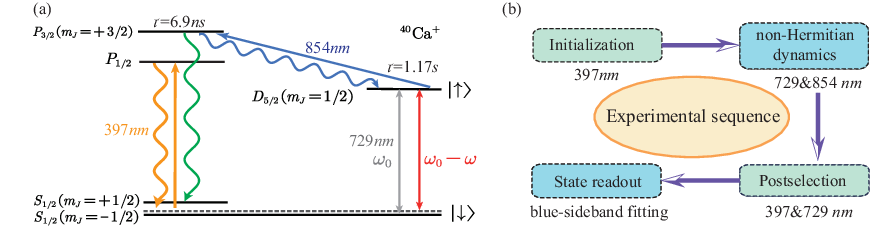}% 
	\caption{\label{A2} Illustration of the experimental protocol. (a) Electronic energy level diagram for $^{40}$Ca$^+$ showing wavelengths associated with relevant transitions. (b) A complete experimental sequence.
	}
\end{figure}	
	
In the following, we examine the mechanism by which a purely dissipative system can effectively simulate a system with balanced gain and loss. By introducing an effective dissipation $-2i\gamma$ on $\lvert \uparrow \rangle$, one could obtain a purely dissipative Hamiltonian  
\begin{equation}
	\begin{aligned}
	\hat{H}_{\mathrm{eff}} &= J_{1}\hat{\sigma}_{x} + J_{2}\left(\hat{a}^{\dagger}\hat{\sigma}_{-} + \hat{a}\hat{\sigma}_{+}\right) - 2i\gamma \ket{ \uparrow} \bra{ \uparrow } \\
	&= J_{1}\hat{\sigma}_{x} + J_{2}\left(\hat{a}^{\dagger}\hat{\sigma}_{-} + \hat{a}\hat{\sigma}_{+}\right) - i\gamma \hat{\sigma}_{z} - i\gamma \hat{I}_{2} \\
	&= \hat{H}_{\mathrm{NH}} - i\gamma \hat{I}_{2}.
	\end{aligned}
	\label{Eq.D3}
\end{equation}
It is shown that, the purely dissipative Hamiltonian $\hat{H}_{\text{eff}}$ and the gain-loss balanced Hamiltonian  $\hat{H}_{\text{NH}}$ differ only by a constant term $-i\gamma \hat{I}_2$, where $ \hat{I}_2$ is the $2 \times 2$ identity matrix. As a result, $\hat{H}_{\text{eff}}$ and $\hat{H}_{\text{NH}}$ share the same eigenstates, and their eigenenergies differ only by a uniform offset $-i\gamma$:  
\begin{equation}
\begin{aligned}		
E_{0}' &= E_{0} - i\gamma, \\ 
E_{n,\pm}' &= E_{n,\pm} - i\gamma \quad (n=0,1,2,\cdots\infty),
\end{aligned}
\label{Eq.D4} 
\end{equation}	
where $\{E_{0}', E_{n,\pm}'\}$ and $\{E_{0}, E_{n,\pm}\}$ are eigenenergies of $\hat{H}_{\mathrm{eff}}$ and $\hat{H}_{\mathrm{NH}}$, respectively.
For the purely dissipative system, the subspace Hamiltonian $\hat{H}_1$ in Eq.~(\ref{eq:Hdirectsum}) of the main text is modified to $\hat{H}_1' = \hat{H}_1 - i \gamma$. Using the original definition of $\mathcal{PT}$ operator, it can be easily shown that
\begin{equation}	
\mathcal{PT}\hat{H}_1' (\mathcal{PT})^{-1} = \mathcal{PT} (\hat{H}_1 - i \gamma)(\mathcal{PT})^{-1} = \hat{H}_1 + i \gamma \neq \hat{H}_1'.
\end{equation}	
Thus, the $\mathcal{PT}$ symmetry is absent in the purely dissipative system. Nevertheless, this has no bearing on the feasibility of using a purely dissipative scheme to simulate a gain-balanced system.
From Eqs.~(\ref{Eq.D3}, \ref{Eq.D4}), it follows that the time-evolution operator of the purely dissipative system is given by $\hat{U}_{\mathrm{eff}}(t) = e^{-\gamma t}\hat{U}_{\mathrm{NH}}(t)$.
Hence, compared with $\hat{H}_{\mathrm{NH}}$, the dynamical evolution of an initial state under $\hat{H}_{\mathrm{eff}}$  
differs only by a global factor $e^{-\gamma t}$. After applying the renormalization procedure, the two systems become completely equivalent.  
This equivalence provides the basis for our proposed experimental scheme, where the desired model can be realized using a purely dissipative system. This is a commonly used method in non-Hermitian quantum systems \cite{PhysRevA.103.L020201,PhysRevLett.126.083604,quinn2023}.

Here, taking the $^{40}\mathrm{Ca}^+$ ion as an example, we provide additional details regarding its experimental implementation, including energy-level selection, control protocols, dissipation engineering, pulse sequences, and measurement methods.

One could consider the qubit states $\ket{\uparrow} \equiv \ket{D_{5/2}(m_J=+1/2)}$ and $\ket{\downarrow} \equiv \ket{S_{1/2}(m_J=-1/2)}$, coupled to the ion’s vibrational mode by sideband transition, as depicted in Fig.~\ref{A2}(a). In the Lamb–Dicke (L-D) regime and under the rotating-wave approximation  (RWA), the simultaneous application of a resonant 729 $nm$ carrier beam and a near-resonant red-sideband beam leads to the Hamiltonian~(\ref{D2}). The purely dissipative Hamiltonian in Eq.~(\ref{Eq.D3}) is realized by coupling $\ket{\uparrow}$ to the short-lived state $P_{3/2}(m_J=+3/2)$ (the lifetime $\tau=6.9\text{\,} ns$) with a $\sigma^+$-polarized 854 $nm$ laser, inducing effective decay from $\ket{\uparrow}$ to $S_{1/2}(m_J=+1/2)$ at a rate $2\gamma$.
	
A complete experimental sequence is illustrated in Fig.~\ref{A2}(b):  
	(i) Initialization: Laser cooling on the 397 $nm$ transition prepares the vibrational ground state $\ket{0}$ \cite{PhysRevResearch.7.013058}. Optical pumping initializes the spin to $\ket{\downarrow}$.  
	(ii) Dynamics: The 729 $nm$ (carrier and red-sideband) and 854 $nm$ lasers implement the coherent and dissipative couplings.  
	(iii) Postselection: Quantum trajectories with unwanted spontaneous jumps (i.e., decays from $P_{3/2}(m_J=+3/2)$ to $S_{1/2}(m_J=+1/2)$) are discarded via consecutive fluorescence detection using the 397 $nm$ laser, assisted by the 729 $nm$ coherent rotations \cite{PhysRevResearch.7.013058}.  
	(iv) Readout: Spin states could be measured by the electron-shelving method \cite{quinn2023,PhysRevResearch.7.013058}. The phonon distribution $P(n)$ can be reconstructed from blue-sideband fitting using a maximum-likelihood method \cite{PhysRevLett.132.130601}. 
	
It is worth noting that the validity of the scheme above relies on the RWA and the L-D approximation. Due to the vibrational frequency of trapped ions which is usually in the order of $MHz$ and much larger than the laser-induced coupling strength, RWA is safely guaranteed even with large excitation number $n$. Consequently, the size of the SSH chain is mainly limited by the L-D approximation. When the site index $n$ becomes large, the L-D approximation $\eta^2(2n+1) \ll 1$ breaks down, and the $\sqrt{n+1}$-dependent intercell hopping relation also becomes invalid \cite{PhysRevA.106.063112}. For $^{40}\mathrm{Ca}^+$ with $\eta \sim 0.05$–0.1, several tens of effective lattice sites are achievable, sufficient to observe the predicted phenomena. With typical parameters ($J_1=2\pi\times200$\,kHz), the maximum evolution time in Fig.~\ref{fig4} corresponds to $0.4$\,ms, well below the coherence times of several milliseconds for both internal and vibrational states \cite{PhysRevLett.117.060504,PhysRevResearch.7.013058}. Although backflow of a small fraction of population from $P_{3/2}$ to $D_{5/2}$ will induce additional dephasing \cite{quinn2023,PhysRevA.109.042205}, this detrimental effect could be small in $^{40}\mathrm{Ca}^+$ due to a preferable branch ratio of spontaneous decays \cite{quinn2023}.

\nocite{*}

\end{document}